\documentclass[journal=jacsat,manuscript=article]{achemso}
\setkeys{acs}{articletitle=true}
\usepackage{amssymb}
\usepackage{float}
\usepackage{graphicx,epsfig,bm,color}
\usepackage{subfig}

\definecolor{NavyBlue}{rgb}{0.00,0.00,0.90}
\definecolor{Red}{rgb}{1.00,0.00,0.00}
\definecolor{Green}{rgb}{0.58,0.0,0.83}

\def\be{\begin{equation}}\def\ee{\end{equation}}
\def\bea{\begin{eqnarray}}\def\eea{\end{eqnarray}}
\def\ba{\begin{array}}\def\ea{\end{array}}

\author{Daniele Varsano}
\affiliation{Center S3, CNR Institute of Nanoscience, Via Campi 213/A, 41125 Modena, Italy}
\author{Giacomo Giorgi}
\affiliation{Dipartimento di Ingegneria Civile e Ambientale, Universit\'a di 
Perugia (DICA), \\
Via G. Duranti, 93 - 06125 - Perugia, Italy}
\author{Koichi Yamashita}
\affiliation{Department of Chemical System Engineering, School of Engineering,\\ The University of Tokyo, 
7-3-1,Hongo, Bunkyo-ku, Tokyo 113-8654 and \\ CREST-JST, 7 Gobancho, Chiyoda-ku, Tokyo 102-0076, Japan
} 
\author{Maurizia Palummo}
\affiliation{Dipartimento di Fisica and INFN, Universit\'a di Roma "Tor Vergata" Via della Ricerca Scientifica 1 Roma, Italy}
\title{
Role of Quantum-confinement in Anatase nanosheets} 

\email{maurizia.palummo@roma2.infn.it, giacomo.giorgi@unipg.it, daniele.varsano@nano.cnr.it}

\begin{document}
\begin{abstract}
Despite most of the applications of anatase nanostructures rely on photo-excited charge processes, yet profound theoretical understanding of fundamental related properties is lacking.  Here, by means of {\it ab-initio} ground and excited-state calculations  we reveal, in an unambiguous way, the role of  quantum confinement effect and of the surface orientation,  on the electronic and optical properties of  anatase nanosheets (NSs). 
The presence of bound excitons extremely localized along the (001) direction, whose existence has been recently proven also in anatase bulk, explains the different optical behavior found for the two orientations when the NS thickness increases. 
We suggest also that the almost two-dimensional nature of these excitons can be related to the improved photo-conversion efficiency observed  when an high percentage of (001) facet is present in anatase nanocrystals. 
\begin{figure}[ht]
\begin{center}
\epsfig{file=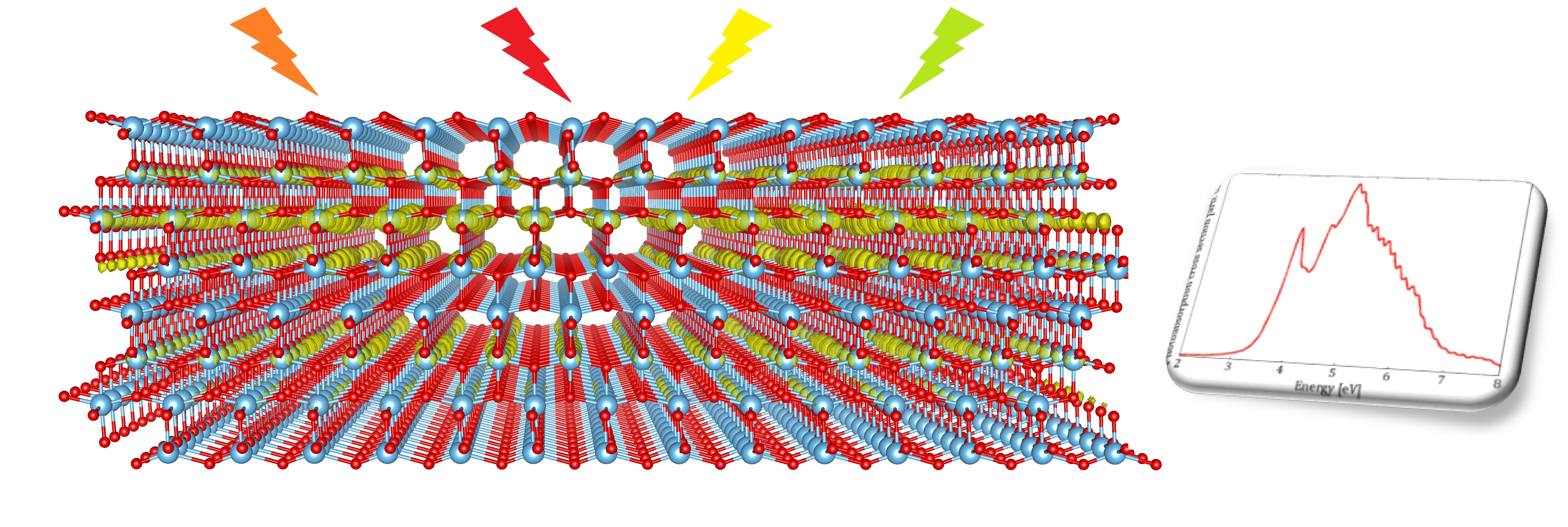,width=0.9\linewidth}
\end{center}
\end{figure}
\end{abstract}

\maketitle
\par 
Among semiconducting oxides, {\rm TiO$_2$} is the most widely used for energy and environmental oriented applications. 
Since 1972 when Fujishima and Honda discovered the phenomenon of photo-catalytic water splitting by shining TiO$_2$ nanoparticles with ultraviolet (UV) radiation,  \cite{FH72} a plethora of works has appeared in literature dedicated to the study of this manifold and extremely appealing material \cite{Fuji99,Diebold2003,Chen2007,Fuji2008}.   
In this regard, nanostructured {\rm TiO$_2$} based materials are largely investigated due to the enhancement of the surface area and to the observed improvement of photo-chemical and photo-physical activity with respect to the bulk phase. \cite{Chen2007,Bavy2009}.
Great attention is devoted to the study of anatase form  which becomes more stable than rutile  at the nanoscale \cite{Gribb1997} and shows superior performances  in photo-voltaic (PV) and in photo-catalytic applications.
Among the possible morphological shapes of anatase nano-materials, the study of (001)-oriented nanosheets (NSs) is becoming particularly attractive in the last years, thanks to several studies which illustrate different routes of  synthesis and  thanks to the indication that, when the (001) facets are the dominant ones, the samples are extraordinary photo-reactive \cite{Gong_JPCB2006,Yang_nature2008,Liu2010,Yang2009, Han2009,Bavy2009,Mogile2008,Vittadini2008,Selloni_2008}. Concerning (101) NSs, some works in the last years have also paid attention to the photo-catalytic activity of anatase nanomaterials with high percentage of these facets. 
Peng et al. \cite{Peng2008} reported excellent photocatalytic activity for samples experimentally obtained via a "{\it chimie-douce}" plus heat treatment method, showing that a large amount of photoactive sites are present, improving the performances with respect to {\it Degussa} samples.
It is important to stress that even if the (001) facets seem to be more photo-reactive than (101) ones, experiments do not provide homogeneous results.  Indeed, in terms of presence of fivefold Ti atoms at the surface, the (001) facet should be more photoreactive than the (101) one, but the opposite should be true because highly  reductive electrons should be generated at the (101), being the  conduction band minimum higher in energy. \cite{Pan2011}

As a partial explanation for the conflicting results concerning (001) and (101) facet photoconversion activity one could adduce the fact that the stabilization of the (001) "bulk--cut" surface, by means of a fluorine mediated passivation technique, is a process that has been only recently theoretically predicted and experimentally developed, \cite{Yang_nature2008,Yang2009} thus providing a conclusive remark about the photoreactivity issue can appear still a quite cumbersome task.
On the other hand, experiments have shown that the (001) surface of anatase TiO$_2$ is much more reactive than the more stable (101) surface. \cite{Xiang2010}
In particular, in the photocatalytic oxidation process the (001) termination is the main source of active sites \cite{Selloni_2008,Yu2010}. Anyway, the presence of less reactive \{101\} facets characterizes the vast majority of anatase nanocrystals. 
According to these findings, it is expected that TiO$_2$ NSs with exposed \{001\} facets have an enhanced photocatalytic activity compared to those nanoparticles where larger is the amount of (101) terminations. \cite{Yang_nature2008,Yu2010}
To further stress the dualism in terms of photoreactivity between the two facets, it is worth pointing out that novel PV technologies based on organic-inorganic halide perovskites \cite{TomTom2009} highly rely on the chemistry of anatase NSs. 
It is interesting to observe that the power conversion efficiency (PCE) of mesoscopic CH$_3$NH$_3$PbI$_3$/TiO$_2$ heterojunction solar cells depends on the TiO$_2$ NS facet exposed. 
When the (001) is the dominant one a PCE of the device double than that obtained with the (101) facet is reported.\cite{LiozEtgar2014}.

Surprisingly,  despite the large number of scientific works,  even on bulk anatase, fundamental properties like the exact value of its electronic gap or the presence of a strongly bound two-dimensional exciton have been only very recently clarified as a result of a combined  experimental and theoretical effort
\cite{Baldini2017}.  
Furthermore the impact of quantum-confinement (QC) effect  on the electronic and optical properties of anatase nanostructures, is still under debate. This is due, on one side, to the fact that experiments reach contrasting results,  being often influenced  by several factors, like synthesis conditions,  presence of defects dopants and co-dopants, \cite{Serpone1995, Reddy2002,Monticone2000}  and on the other hand  to the lack of results obtained by means of predictive quantum-mechanical excited-state calculations.
Then our goal here is to investigate the role played by QC effect, focusing  
both on the electronic and optical properties of (001) anatase NSs of increasing thickness and using the (101) NSs to point out also the role of different orientations.

For this reason, we have specifically selected NSs with the simplest surface models, avoiding reconstructed and/or hydroxylated facets, with the idea to leave to future investigations the study of many-body effects in NSs with less ideal surface motifs. 
By means of Density Functional Theory (DFT) and post-DFT excited-state (namely GW ad Bethe-Salpeter Equation(BSE)) calculations, we demonstrate the mismatch between electronic and optical gap, with the latter mainly associated to  the presence of strongly bound bidimensional excitons, confirming the results recently reported by Baldini et al. \cite{Baldini2017} for the bulk anatase.

\begin{figure}[htp]
  \centering
\includegraphics[width=\textwidth]{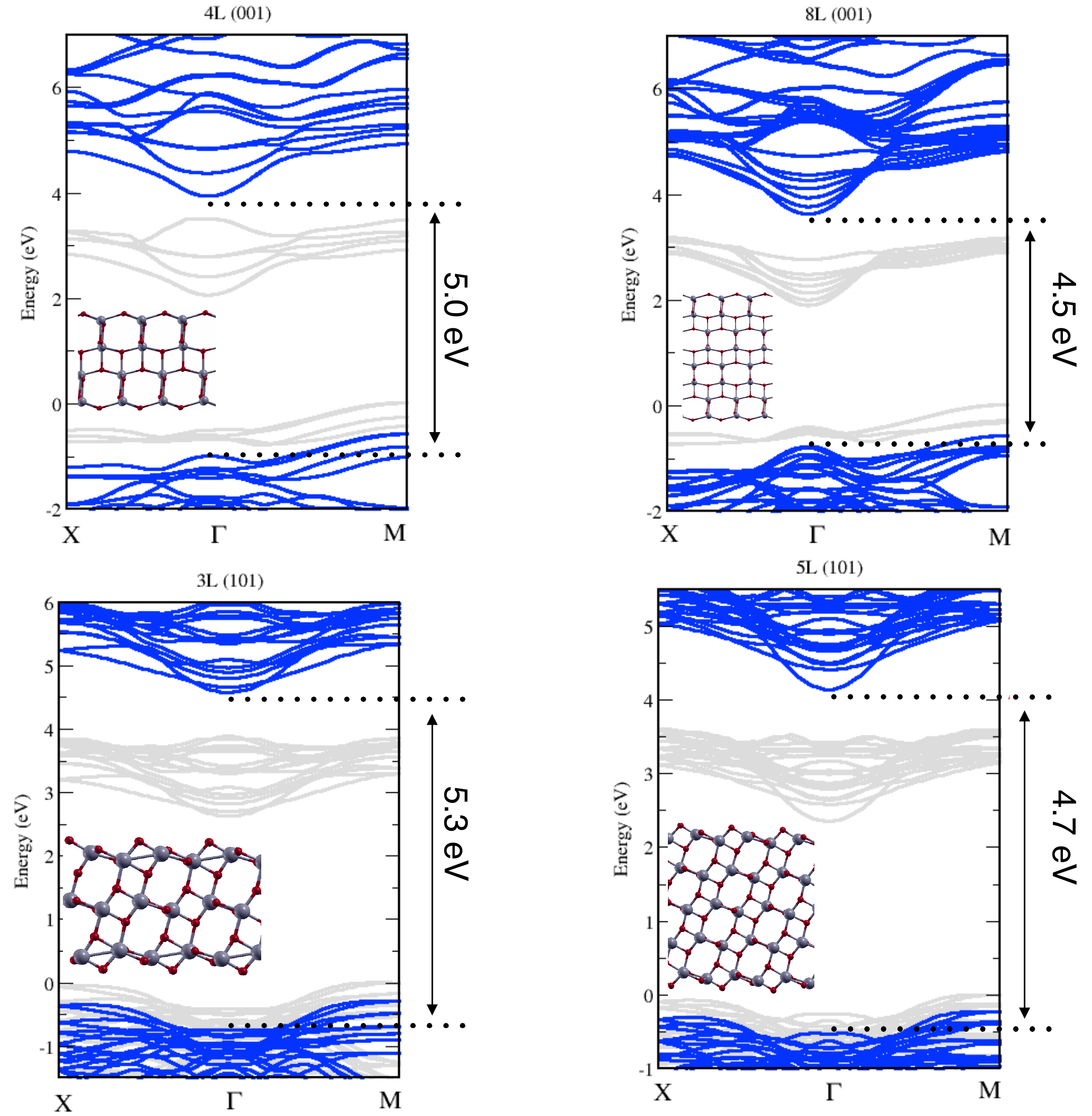}
\caption{Quasi-Particle bandstructures (blue lines) of the 4L, and 8L (001) and 3L and 5L (101) NSs. To show the entity and the small {\it k}-dependence  of the QP correction some PBE conduction and valence bands are also reported (gray). The QP corrections were calculated on a regular grid and band interpolation was performed following the method of Ref.~\cite{agapito2013effective} using the WanT package\cite{want}}
\label{Fig1}
\end{figure}

\begin{figure}[ht]
\begin{center}
\includegraphics[width=\textwidth]{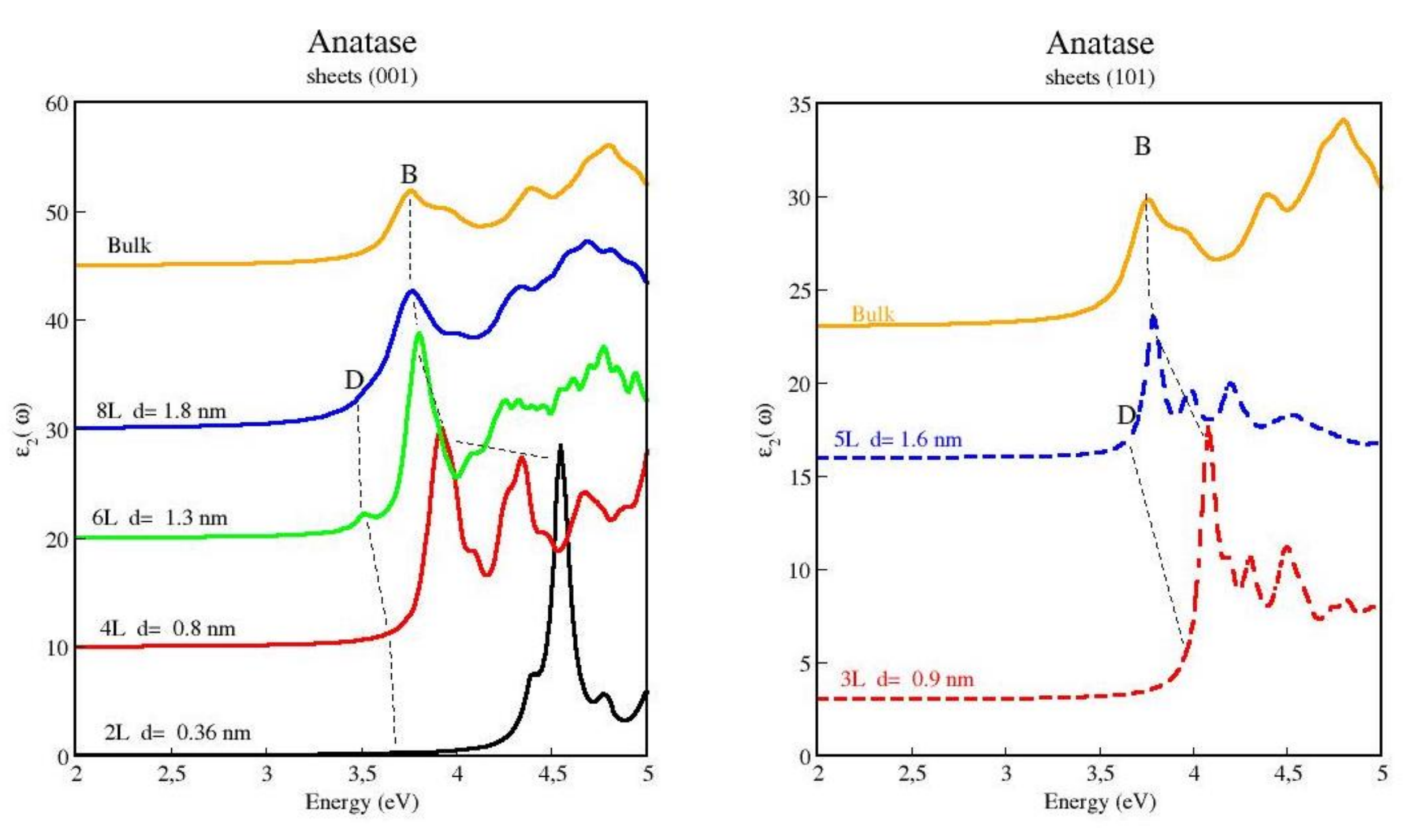}
\end{center}
\caption{Optical spectra of bulk, anatase (001) and (101)
for light polarized $\perp$ {\it c}-axis.
First Dark (D) and Bright (B) exciton are also indicated.
For NSs the intensities are renormalized to their effective thickness.}
\label{Fig2}
\end{figure}

Ground-state atomic structures have been relaxed using DFT\cite{DFT,KS},
as implemented in the VASP package\cite{VASP}, within the generalized gradient approximation (GGA) of Perdew-Burke-Ernzerhof (PBE).\cite{PBE}
The Bl\"ochl all-electron projector augmented wave (PAW) method~\cite{PAW} has been employed. For Ti atoms a 12{\it e} potential has been employed along with a cut-off energy set to 500.0~eV for the plane wave basis.
A force convergence criterion of 0.035~eV/\AA~has been chosen, while different samplings Gamma centered of the Brillouin zone have been employed according to the lateral dimension of the system under investigation.
Using the relaxed atomic structures we performed further relaxation of the atomic structures using the Quantum Espresso package  \cite{PWSCF} in a consistent PBE scheme but using norm-conserving pseudopotentials  with a plane-wave expansion of 2450 eV of cutoff.
Then self-consistent and non self-consistent calculations have been performed to obtain DFT-KS eigenvalues and eigenvector to calculate the quasi-particle (QP) energies in GW approximation and optical excitation energies solving the Bethe-Salpeter equation (BSE)\cite{Hanke74,Hanke80,Strinati82,Strinati88} by means of the many-body code YAMBO \cite{yamboproject}.
For the GW simulations a plasmon-pole approximation for the inverse dielectric matrix has been applied, \cite{Godby1} 136 eV (952 eV) are used for the  correlation $\Sigma_{c}$ (exchange $\Sigma_{x}$) part of the self-energy and the sum over the unoccupied states for $\Sigma_{c}$ and the dielectric matrix
is done up to about $\sim$50~eV above the VBM. In order to speed up the convergence with respect empty states we adopted the technique described in Ref.~\cite{Bruneval08}.
Finally when the QP energies and eigenfunctions are known, the optical properties are calculated
by solving the Bethe-Salpeter equation (BSE) where the electron-hole interaction is also taken into account
~\cite{RevOnida,review-rohlfing,yamboproject}.
A {\it k}-points grid of $10\times10\times1$ ($10\times5\times1$) has been used in the GW and BSE calculations for (001) [(101)] sheets.
A cutoff in the coulomb potential in the direction perpendicular to the 
sheet, 
has been used in the excited state calculations to eliminate the spurious interactions along the non periodic direction and to simulate a real isolated nanosheet~\cite{rozzi06}
.  
Since our goal is mainly to analyze the role of quantum confinement
we focus first on four (001)-oriented anatase NSs of increasing thickness. Then   
in order to capture the role of the surface orientation, we perform  simulations also on two (101)-oriented NSs of different thickness.
For the (001) we use a (1$\times$1)  cell with the lattice parameter equal to the bulk one, allowing a full relaxation of the atomic positions.  We adopt this choice because we consider the experimental conditions of the delamination process along [001] direction of anatase \cite{Mogile2008} and we aim to compare the results with the corresponding bulk data.  
Two asymmetric Ti-O bonds (1.74 and 2.24~\AA) at surface are formed. In a previous work\cite{Palummo2012}, where we focused on anatase and lepidocrocite thinnest sheets, we have named this structure Anatase-AS2.
Despite the asymmetric Ti-O bond structure has not been detected experimentally, the larger stability of the asymmetric bond type with respect to bulk-terminated symmetric one is widely predicted at the theoretical level \cite{Lazzeri01}.
This result can be ascribed to the residual stress amount present in the reconstruction.
We focus on (001)-oriented NSs formed by two, four, six,  and eight atomic layers with an estimated atomic thickness of 0.36, 0.80, 1.3, and 1.8 nm, respectively.
The lateral views of two of these atomic structures are shown in the insets of Fig.\ref{Fig1}.
\noindent
Concerning the (101) orientation we assemble two NSs with thickness of about 0.95 nm and 1.65 nm, respectively  and using a 1$\times$1 cell in the surface plane.  
Similarly to the (001) sheets, we have kept the in-plane lattice parameters frozen to the bulk optimized value,  i.e. 3.78 $\times$ 10.24 ~\AA, to focus only on the confinement effects. 
The QP bandstructures of two (001) and  two (101) oriented NSs (blue curves) of different thickness, plotted 
along the high-symmetry directions X$\rightarrow \Gamma\rightarrow$M (where X = (0.5,0,0), $\Gamma$=(0,0,0) and M= (0.5,0.5,0) in reciprocal lattice units), 
are shown in Fig.  \ref{Fig1}.  
Few conduction and valence bands around the Fermi energy,  calculated at the DFT-KS level, are also reported (gray lines).    
First of all we can observe  that the self-energy correction to the KS gap increases reducing the sheet thickness and is larger than the value found in anatase bulk which is of the order of 1.4 eV (see ref. \cite{Baldini2017} for more details). This is consistent with the fact that the dielectric screening reduces decreasing the size of the nanostructure\cite{bruno_prl2007}.
Moreover, at a given thickness, a small but not negligible {\it k}-dependence of the QP correction is found and for this reason we avoid the use of a rigid scissor operator to open the unoccupied bands.  
We finally point out that the indirect gap character typical of anatase bulk remains in all the considered NSs, both at DFT and QP level of approximation. 
For a comparison with anatase bulk band structure, obtained at the same level of theoretical approximation, we refer the reader to Fig.5 of ref.\cite{Baldini2017}. 
Fig.\ref{Fig2}(a) shows the optical spectra of the (001) anatase nanosheets 
(for light-polarized $\perp$ to the {\it c}-axis) compared to the corresponding optical spectrum of anatase bulk. Self-energy, local-fields, and excitonic effects are included through the solution of the Bethe-Salpeter equation. 
It is important to underline that due to the depolarization effect \cite{Ajiki1994,BrunoSS2007,Palummo2012} the optical spectra for light
depolarized perpendicularly to the nanosheet (not reported here) are almost zero when, as in the present study, the single-particle
approach is overcome taking into account local-field effects.
First of all we note that increasing the sheet thickness, the position of the first optical peak rapidly recovers the bulk-like position. Moreover, from the analysis of the optical spectrum in terms of excitonic eigenvalues and eigenvectors, we know that, while in the bulk the first optical peak (at $\sim$3.8 eV) corresponds to a bright exciton (indicated by B in the figure) and any lower dark exciton is  present, in the NSs several excitons with weak oscillator strength 
appear, where the position of the first one is indicated in the figure by D. 
Fig.\ref{Fig2}(b) shows the corresponding optical spectra calculated for the two (101)-oriented NSs.
In this case a larger QC effect with respect to the other orientation is clearly visible. 
As we will discuss later on, this different behavior due to quantum-confinement 
is strictly related to the spatial character of the first exciton.
\noindent

\begin{figure}[ht]
\begin{center}
\epsfig{file=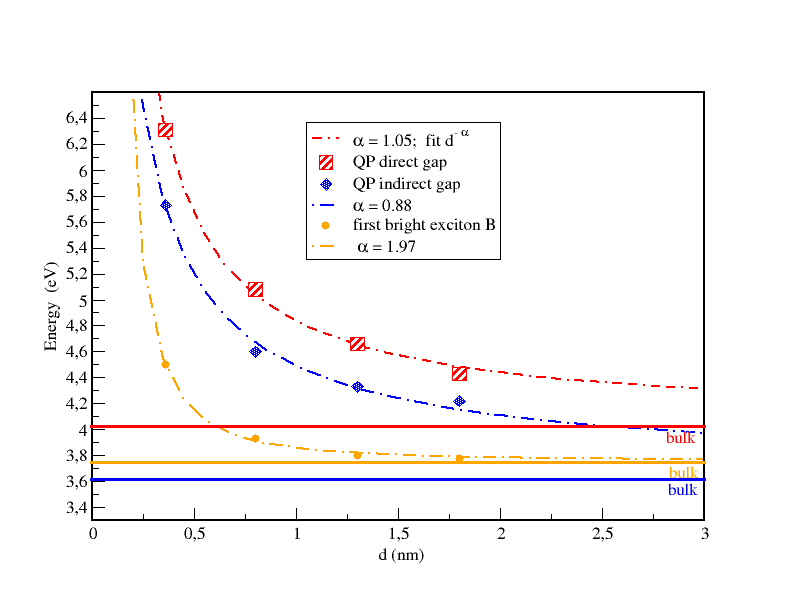,width=0.7\linewidth}
\end{center}
\caption{Direct (squares) and Indirect (diamonds) electronic (QP) gaps; first bright (circles) excitons  (BSE) as function of NS thickness. Scaling law fits $\simeq 1/d^\alpha$, where $d$ is the NS thickness, are similarly reported. The red (blue) and yellow solid lines represent the QP direct (indirect) and optical direct gap in bulk.
\label{Fig3}}
\end{figure}

To further illustrate the role of quantum-confinement in Fig.\ref{Fig3}  we report the value of the direct and indirect QP electronic gaps together with the energy of the first bright  exciton as function of the (001) sheet thickness; 
since we considered only two (101) sheets of different thickness we did not include the corresponding data for this orientation.

As reported in previous literature (see i.e.  refs. \cite{bruno_prl2007,yan2007,delerue2000}) focusing on the size dependence of the electronic and optical gaps in semiconducting nanostructures, the direct QP electronic gap values are fitted with the scaling law $E_{gap}^{bulk} + C/d^{\alpha}$, where $E_{gap}^{bulk}$ is the bulk gap value and $d$ is the NS thickness.  Similar scaling laws have been used to fit the indirect QP electronic gaps and the energies of the first bright excitons (B).

Still looking at Fig.\ref{Fig3}, it is clear that the electronic and optical gaps have a different behavior decreasing the thickness of the nanosheet.
Indeed, the electronic QP direct and indirect gaps remain larger than the corresponding bulk value  for the considered thicknesses. 
This can be explained by the fact that the self-energy correction, due to the reduced dielectric screening and localization of the wavefunctions, strongly increases reducing the NS thickness. The scaling exponents able to fit the direct and indirect gaps are 1.05 and 0.88, respectively. Similar exponents ($\sim$1) have been obtained in previous works \cite{yan2007,bruno_prl2007,delerue2000} that take into account the many-body self-energy corrections in nanostructures of different dimensionality. 

On the other hand, as the NS thickness increases, the optical (excitonic) direct gaps converge more rapidly to the corresponding optical direct gap of the bulk.
Again this finding is consistent with previous studies of many-body effects in low-dimensional materials \cite{delerue2000,bruno_prl2007,Palummo2012} 
and can be explained in terms of compensation of the induced polarization effect present both in the quasi-particle self-energy and in the Bethe-Salpeter kernel. In other words, for the optical data, the convergence is reached as soon as the bulk-like excitonic wave function is contained in the NS thickness which is at $d \sim$ 2 nm \cite{varsano_prl2008}. 
As already observed in other studies,\cite{bruno_prl2007,delerue2000} the scaling exponent able to fit the optical data results larger (for the anatase nanosheets is 1.97) than the value used to fit the electronic QP gaps. 
It is worth observing that although this value is very similar to 2 $‒-$ the exponent of the ideal particle-in-a-box model $‒-$ the physics of the exciton described here is completely different and this value can be understood only in terms of the cancellation of the induced polarization effect which is present both in the GW and in the BSE kernel that then rapidly cancels out as the size of the nanostructure increases.


We aim now to discuss the spatial character of the first bright exciton (B) of the (001)- and (101)-oriented NSs and to compare it with the corresponding character recently found by Baldini et al. in anatase bulk \cite{Baldini2017}. 
Indeed, the direct optical gap of single crystal anatase is dominated by a strongly bound exciton, rising over the continuum of indirect interband transitions, which possesses an intermediate character between the Wannier-Mott and Frenkel regimes  and  displays  a  peculiar 2D wavefunction in  the 3D lattice.

\begin{figure}[ht]
\begin{center}
\includegraphics[width=0.7\linewidth]{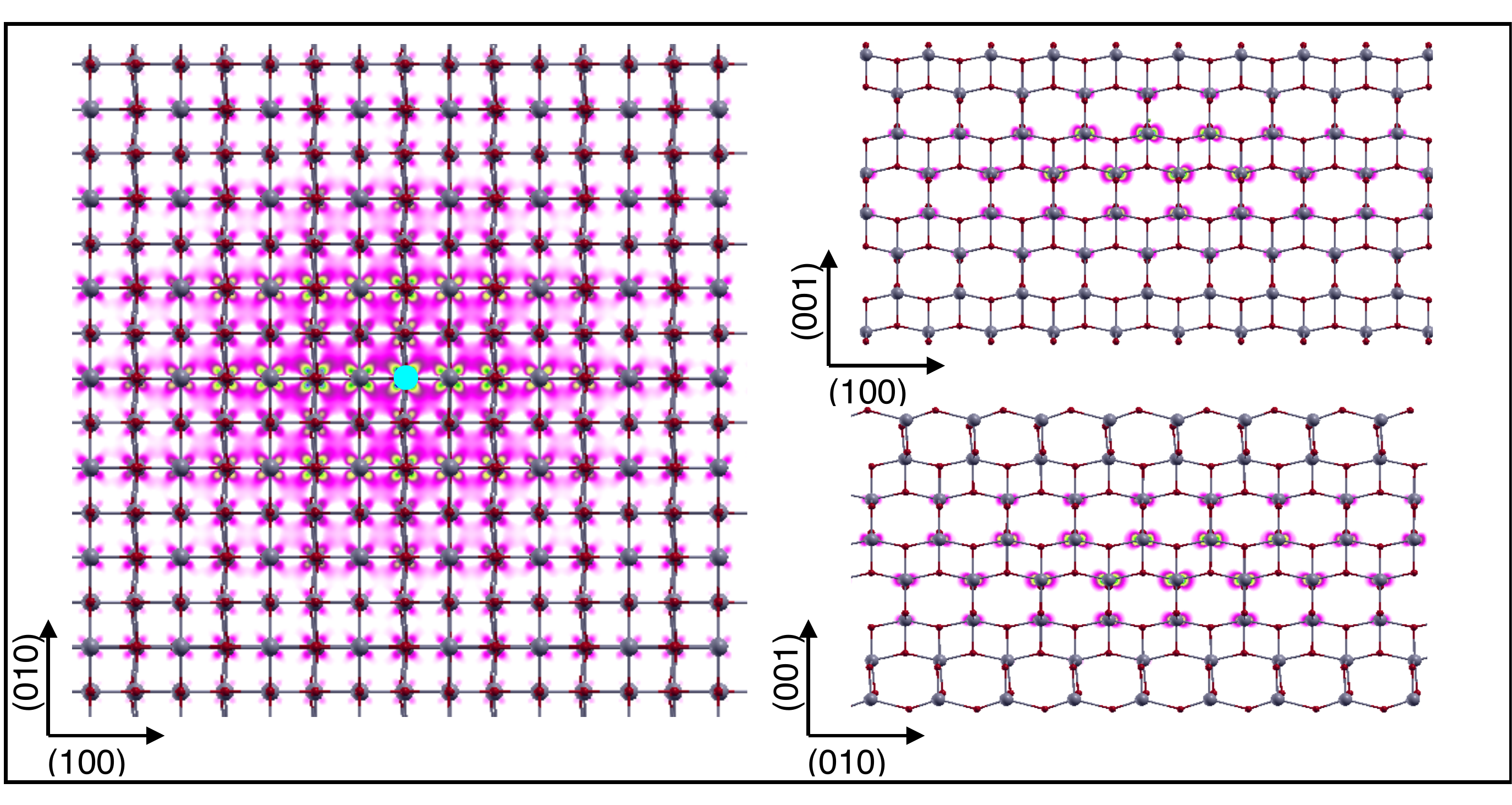}
\includegraphics[width=0.7\linewidth]{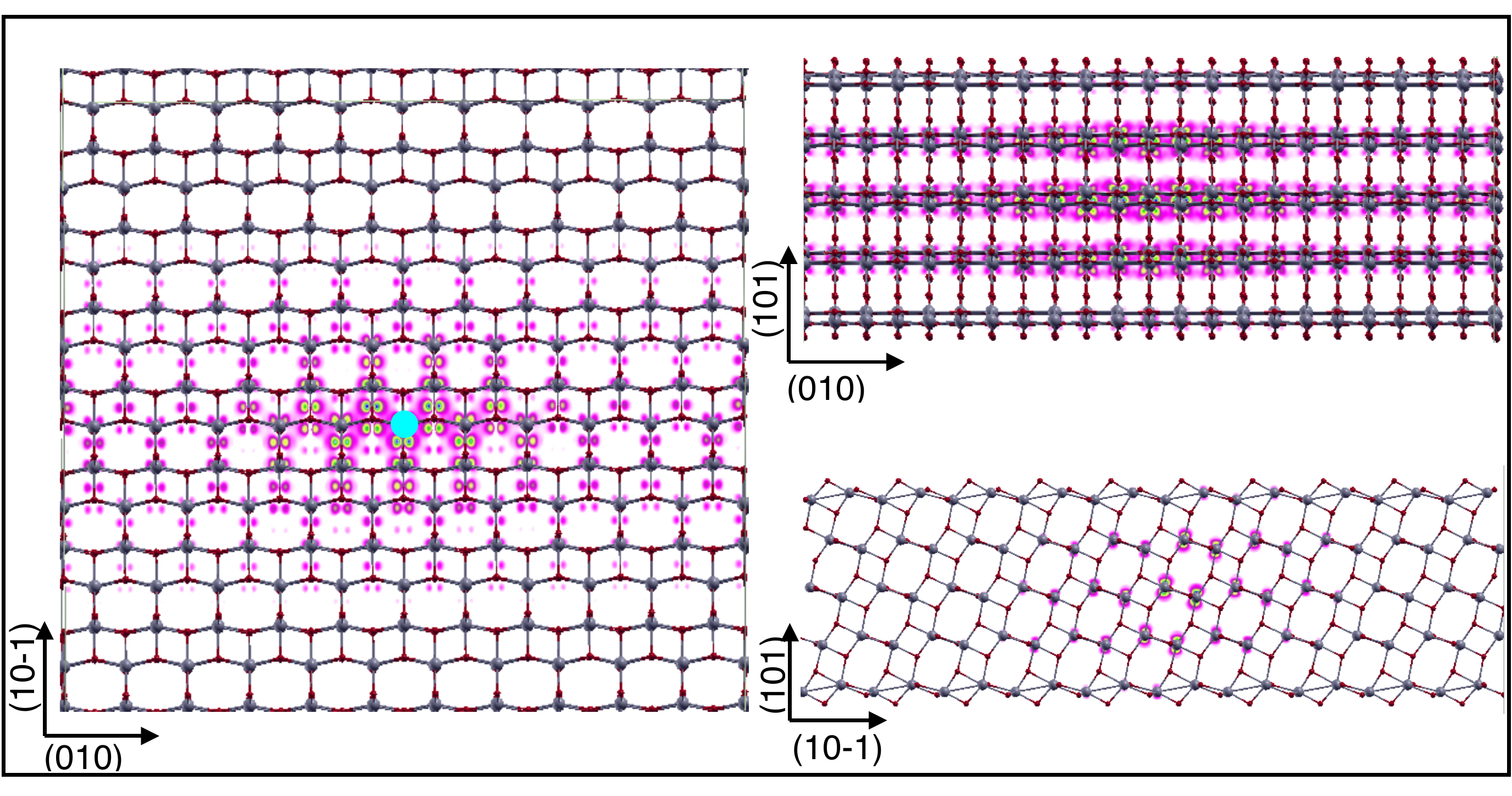}
\end{center}
\vskip 1.5 truecm
\caption{ Top (left) and side (right) views of the 
spatial distribution of the first exciton for the thicker (001) (top panel) and (101) (bottom panel) sheets. The light-blue ball represents the hole position.
\label{Fig4}}
\end{figure}

By fitting the exciton wavefunction with a 2D hydrogen model, the exciton Bohr radius  results $\sim$ 3.2 nm,  while the 90\% of the excitonic squared modulus wavefunction is contained within 1.5 nm\cite{Baldini2017}. 
The top panel of Fig.\ref{Fig4} reports the exciton spatial distribution for the (001)-NS ($d= 1.8$ nm), showing the same high degree of spatial localization, along the c-axis, observed in bulk (see ref.\cite{Baldini2017}), while the bottom panel reports the corresponding distribution for the (101) sheet of thickness $d=1.6$ nm.
In both cases, fixing the hole position (blue-light ball in the figure) in the proximity of a bond that an oxygen atom forms with a titanium one, the probability to find the photo-excited electron in or near the (001) crystallographic plane where the hole has been created is large, while it rapidly decreases along the [001] direction.

It is worth pointing out that this localized behavior is substantially the same  placing the hole in different positions and that it holds both 
in the case the hole is fixed in a bulk (shown in Fig.4) or in a sub-surface position (not shown here). 
The analysis of the dark states (not shown here) does not provide noticeable differences in the spatial exciton localization plots with respect to the bright exciton case. 
Furthermore, from the corresponding analysis of the (001) nanosheet we have found that the B-exciton is composed by the mixing of single-particle vertical transitions mainly at $\Gamma$ (with smaller contribution from points near it)  between VBM-1 to CBM+1 and VBM-4 to CBM, while for the D-exciton the involved transition is that from VBM to CBM, still at $\Gamma$.  From the spatial analysis of these states they are mainly localized in the central part of the sheet.
Minor contribution from other points near $\Gamma$ and other bands like VBM-3 to CBM,CBM+1 and VBM-2 to CBM,CBM+1 occurs.
This finding is consistent with what observed in bulk, where the excitonic wavefunction of the first bright exciton results formed by the mixing of transitions from VBM ,VBM-1 to CBM along the $\Gamma$$\rightarrow Z$  direction (where the bands near the gap are almost parallel, see Fig. 5 of ref.\cite{Baldini2017}).
Indeed, it is important to recall that all the points along the bulk $\Gamma$$\rightarrow Z$ direction are folded in $\Gamma$ in the (001) nanosheet. 

Then from this specific, quasi 2D, spatial distribution of the exciton, mainly induced by the lattice geometry  (see ref.\cite{Baldini2017} for more details),  we can deduce some conclusions.
Due to the extreme localization of the exciton along the [001] direction, when the hole is created in a bulk position the exciton 
does not touch the surfaces already for thickness of the (001) NS of the order (or larger) of 1.8 nm. As consequence, as we have shown in Fig.\ref{Fig2},  a  bulk-like behavior of the optical spectrum  is rapidly recovered in the (001)-oriented NSs,  while a larger QC effect is visible  in the (101) NSs of comparable thickness, due to the fact that in this case  the excitonic wavefunction extends up to the two surfaces, remaining confined.

Moreover, we suggest that it could contribute to the larger photo-reactivity  often reported in anatase nanostructures when a large percentage of (001) facets is present. Indeed recent experiments have shown that i) in the photo-catalytic process the reduction and oxidation reactions preferably occur on (100)/(101) and (001) facets,  respectively \cite{Kim_prb2017} ii) the thickness of anatase nanocrystals when a large percentage of (001) facets is present, is thinner in the [001] than in other crystallographic directions. \cite{Yang_CEC2011}
These two facts, in addition to the observed exciton spatial distribution, suggest that when a large percentage of (001) facet is present,  
the photo-excited hole can easily reach the (001) termination, especially if created not far from it  and, at the same time, due to the delocalized  nature of the exciton in the (001) plane, there is a non-zero probability to collect instantaneously the electron  at other terminations, like the (101) surfaces.
Although at the moment this is only a speculation, we point out that recent works in organic \cite{Tamura2013,Caruso_pnas2012} and hybrid organic-inorganic \cite{Ahmad_2015} PV materials show that the presence of extremely delocalized coherent excitonic bound states, due to the low dielectric screening, can contribute to enhance the photo-conversion efficiency. \\
To summarize, by means of the Bethe-Salpeter equation solution, we have here investigated the excitonic behavior of anatase nanosheets with majority and minority surface orientation. In particular, for (101) and (001)-oriented NSs, we have focused on quantum-confinement effects and on the role they play on the optical properties of such NSs. While  the former show a more marked QC ascribed to the exciton confinement induced by the two \{101\} delimiting surfaces, in the case of the (001) orientation we instead observe the existence of a threshold thickness value ($\sim$ 2 nm) above which the bulk optical behavior is recovered.
The specific quasi two-dimensional character of the exciton can  be related to the difference in terms of photo-reactivity between the two orientations, with relevant consequences in devices exposing the two different facets.

\section{{\bf Acknowledgments}}  
The authors acknowledge Annabella Selloni for the useful discussions and for kindly reading the manuscript.
M.P. acknowledges INFN for financial support through the National project {\rm Nemesys} and for allocated 
computational resources at Cineca with the project {\rm INF16-nemesys} and the EC for the RISE Project No. CoExAN  GA644076. 
D.V. thanks Andrea Ferretti for permitting the use of the developer's version of the WanT package and acknowledges partial support from the EU Centre of Excellence “MaX-Materials Design at the Exascale” (Grant No. 676598). 
We acknowledge PRACE for awarding us access to resource Fermi based in Italy at CINECA (Grant No. Pra12\_3100 and Pra14\_3664 ). G.G. wants to thank CINECA (ISCRA project ref. HP10C79G0F).

\bibliography{anat}
\end{document}